\documentclass{article}

\usepackage[english]{babel}
\usepackage[letterpaper,top=1.8cm,bottom=1.8cm,left=1.8cm,right=1.8cm]{geometry}
\usepackage{newtxtext,newtxmath}
\usepackage[final]{microtype}
\usepackage{amsmath}
\usepackage{graphicx}
\usepackage{subcaption}
\usepackage[colorlinks=true, allcolors=blue]{hyperref}
\usepackage{float}
\usepackage{xcolor}
\usepackage{wrapfig} 
\usepackage{listings}
\usepackage{booktabs}
\usepackage{hyperref}
\usepackage{authblk}

\title{Monosemanticity in Recommender Systems}
\author[1]{Yagel Alfasi}
\author[1]{Eden Rzezak}
\author[1]{Eadan Schechter}
\affil[1]{School of Industrial \& Intelligent Systems Engineering, Tel Aviv University}
\date{\today}

\begin{document}
\maketitle

\begin{abstract}
Latent factor models such as matrix factorization are widely used in recommender systems, yet the learned embedding dimensions typically lack explicit semantic interpretation. This opacity limits transparency, explainability, and principled intervention in recommendation behavior. While sparse autoencoders (SAEs) have recently been used to extract monosemantic features from dense neural representations, standard SAEs suffer from scaling pathologies including feature splitting, feature absorption, and feature composition, which degrade interpretability as dictionary size increases.

In this work, we investigate whether hierarchical sparse representations can reveal interpretable structure in collaborative filtering embeddings. We train a large-scale matrix factorization recommender system on the Amazon Fashion dataset and apply a Matryoshka Sparse Autoencoder (MSAE) to the learned embeddings. We analyze the resulting latent features through metadata alignment and LLM-generated labeling to assess semantic coherence and disentanglement. Finally, we show an intervention on a subset of gender associated latent neurons that emerged from the analysis. Our findings suggest that collaborative filtering embeddings contain recoverable hierarchical structure, and that Matryoshka training provides a principled mechanism for exposing interpretable latent factors in interaction-driven recommendation models.

The full Git repository is available at 
\href{https://github.com/YagelAL/MonosemanticityRS}{Project GitHub Repository}.
\end{abstract}

\section{Introduction}
Recommender systems play a central role in modern digital platforms, shaping how users discover content in domains such as movies, music, and e-commerce \cite{Ricci2011IntroRecSys}. Among the most influential approaches are latent factor models, particularly matrix factorization (MF), which represent users and items as dense embedding vectors learned  through their interactions to estimate relevance \cite{Koren2009}. Despite their empirical success, these latent representations remain opaque and difficult to interpret, as individual embedding dimensions rarely correspond to clear, coherent, human-interpretable semantic concepts.\cite{Zhang2018Explainable}.

This lack of interpretability poses several challenges, especially in high-stakes domains such as e-commerce, where understanding why a product is recommended is crucial for transparency, auditing, and user trust \cite{Zhang2018Explainable}. From a practical standpoint, it limits the ability to debug models, understand failure cases, or reason about bias and fairness in recommendation outcomes \cite{Abdollahi2019ExplainableMF,Burke2020FairnessRecSys}. From a scientific perspective, it obscures how recommender systems internally represent user preferences and content structure \cite{DoshiVelez2017Interpretability}. As recommender systems increasingly influence real-world decisions, the need for transparent and interpretable representations becomes more pressing. 

Recent advances in mechanistic interpretability suggest that dense neural representations may conceal underlying sparse, semantically coherent structure. Sparse autoencoders (SAEs) have been successfully applied to language models to extract monosemantic features - latent units aligned with single human-interpretable concepts \cite{Olah2020Circuits,Anthropic2024Scaling}. However, scaling standard SAEs introduces critical failure modes including feature splitting, feature absorption, and feature composition, which degrade interpretability despite improved reconstruction performance.

Motivated by these advances, this project explores whether similar ideas can be applied to recommender systems. Unlike language models, recommender systems rely fundamentally on interactions between user and item embeddings, rather than on a single shared representation space \cite{He2017NCF}. This raises a natural question: \emph{can sparse autoencoders extract monosemantic, interpretable factors from recommender embeddings trained on interaction data?}

Matryoshka Sparse Autoencoders (MSAEs) were recently proposed as a hierarchical alternative that mitigates the SAEs pathologies by simultaneously training multiple nested dictionaries of increasing size \cite{Bussmann2025Matryoshka}. Instead of optimizing a single flat sparse representation, Matryoshka SAEs enforce reconstruction at multiple prefix sizes, encouraging early latents to capture general concepts and later latents to refine them. This nested training objective prevents later specialized features from absorbing or fragmenting earlier general ones.

This hierarchical structure is particularly well-suited to recommender systems. Product domains such as fashion naturally exhibit multi-level semantic organization (e.g., clothing → women’s clothing → dresses → summer dresses → floral summer dresses). If collaborative filtering embeddings encode such hierarchical structure implicitly, then a flat SAE may distort it, whereas MSAE may recover it.

In this work, we investigate whether matrix factorization embeddings trained on the Amazon Fashion dataset contain recoverable hierarchical monosemantic structure. We first train a large-scale MF recommender system using sparse interaction matrices. We then apply a Matryoshka Sparse Autoencoder to the learned item embeddings in a post-hoc interpretability framework. Finally, we label latent neurons to assess semantic coherence and disentanglement and show a gender specific intervention.

\paragraph{Contributions}
This work makes the following concrete contributions:

\begin{itemize}
    \item \textbf{Application of hierarchical sparse autoencoders to recommender systems.}
    We introduce a post-hoc interpretability pipeline that applies Matryoshka Sparse Autoencoders (MSAEs) to matrix factorization embeddings trained on large-scale e-commerce interaction data.  
    \item \textbf{Adaptation of the monosemanticity score to collaborative filtering embeddings.}
    We adapt the monosemanticity metric proposed by Pach et al. (2025)\cite{Pach} to recommender system representations by replacing image similarity with cosine similarity between items in the learned matrix factorization embedding space, enabling quantitative evaluation of semantic coherence in recommender embeddings.   
    \item \textbf{Empirical evidence of hierarchical semantic structure in recommender embeddings.}
    Through quantitative analysis and qualitative neuron labeling, we show that MSAE latent neurons organize along a semantic gradient, where early neurons capture broad product categories while later neurons specialize in finer-grained concepts on the Amazon Fashion dataset.   
    \item \textbf{Discovery and causal validation of a gender-sensitive latent axis.}
    We identify a subset of latent neurons associated with women-oriented product categories and demonstrate through controlled activation interventions that manipulating these neurons causally shifts the gender composition of recommendation lists while maintaining high overlap with the original recommendations.    
    \item \textbf{Empirical comparison between SAE and MSAE representations.}
    We show that imposing hierarchical sparsity with MSAE preserves the ranking behavior of the underlying MF recommender relative to a standard SAE baseline enabling a structured interpretability analysis framework.
\end{itemize}

By bridging mechanistic interpretability and recommender systems, this work examines whether hierarchical sparse representations can expose structured semantic factors in interaction-driven embeddings. This work complements recent efforts to extract monosemantic structure from recommender embeddings \cite{WWW2026MonosemanticRecSys}.

\section{Related Work}
\subsection{Interpretability in Recommender Systems}
Interpretability has long been a challenge in recommender systems based on latent representations \cite{Zhang2018Explainable}. Early approaches focused on explicitly interpretable models, such as neighborhood-based methods or factor models with constrained semantics \cite{Koren2009}. However, these models often sacrifice predictive performance compared to unconstrained latent factor approaches.

More recent work has explored post-hoc interpretability techniques, including feature attribution, perturbation-based explanations, and attention mechanisms \cite{NAIS}. While such methods provide insights into individual predictions, they typically do not reveal the semantic structure of the latent embedding space itself \cite{Abdollahi2019ExplainableMF}. As a result, the internal representations learned by matrix factorization remain largely opaque.

\subsection{Sparse Autoencoders}
Sparse autoencoders (SAEs) are neural architectures designed to learn representations in which only a small subset of latent units are active for any given input. By combining a reconstruction objective with explicit sparsity constraints, typically enforced via $\ell_1$ regularization or KL-divergence penalties, SAEs encourage compact and selective latent representations that differ fundamentally from dense embedding spaces \cite{Ng2011SparseAE}.

Recent work in mechanistic interpretability suggests that sparsity can play a crucial role in mitigating feature superposition, motivating the use of sparse autoencoders as a tool for extracting more interpretable latent representations \cite{Anthropic2024Scaling}.

\subsection{Monosemanticity and Mechanistic Interpretability}
Mechanistic interpretability seeks to explain neural network behavior by identifying meaningful internal components, such as neurons, and understanding how they contribute to model outputs \cite{Olah2020Circuits}. A central challenge is that modern neural networks often rely on highly distributed representations, in which individual neurons or dimensions participate in encoding multiple unrelated concepts. This phenomenon, commonly referred to as \emph{feature superposition}, complicates attempts to assign clear semantic roles to internal units.

Within this framework, monosemanticity has emerged as a desirable property of representations: a monosemantic feature corresponds to a single, coherent, human-interpretable concept and exhibits consistent behavior across contexts. Rather than assuming that neural networks are inherently monosemantic, recent interpretability research frames monosemanticity as something that can be induced through architectural and training choices. In particular, imposing sparsity has been proposed as a mechanism for reducing superposition by encouraging models to allocate separate representational capacity to distinct features.

Empirical support for this perspective has been provided by recent large scale studies demonstrating that sparse autoencoders trained on neural activations can recover features that align with coherent semantic concepts in large language models \cite{Anthropic2024Scaling}. These results suggest that dense representations may obscure an underlying, more interpretable structure, and that sparsity can serve as a lens through which this structure becomes accessible.

While the majority of monosemanticity research has focused on language and vision models, its extension to recommender systems remains relatively unexplored. Recommender models differ fundamentally in that semantic meaning arises from interactions between user and item embeddings, rather than from a single shared activation space. A recent study addresses this challenge by proposing prediction-aware sparse autoencoders that preserve user-item interaction semantics while extracting monosemantic features from recommender embeddings \cite{WWW2026MonosemanticRecSys}. 

This project complements that line of work by examining a classic matrix factorization setting and focusing on exploratory analysis of sparsity induced semantic structure and features. By situating monosemanticity within the broader mechanistic interpretability framework, we aim to better understand when and how interpretable latent factors can emerge in interaction-based recommendation models.

\subsection{Matryoshka Sparse Autoencoders}
Standard sparse autoencoders optimize a single reconstruction objective under sparsity constraints. While effective at inducing selective activations, scaling the dictionary size often leads to feature splitting, feature absorption, and feature composition, degrading interpretability \cite{Chanin2024}.

Matryoshka Sparse Autoencoders (MSAEs) address this limitation by training multiple nested reconstruction objectives simultaneously \cite{Bussmann2025Matryoshka}. 

Denote $x \in \mathbb{R}^d$ as an input embedding (in our setting, item or user embeddings from matrix factorization). An encoder produces a sparse latent vector:

\[
z = h_{enc}(x) \in \mathbb{R}^k.
\]

Given a set of ordered prefix sizes
\[
\mathcal{M} = \{m_1 < m_2 < \dots < m_L = k\},
\]

each prefix $z^{(m)} = (z_1, \dots, z_m)$ is independently decoded using only the corresponding subset of decoder weights:

\[
\hat{x}^{(m)} = W_{dec}^{0:m} z^{(m)} + b.
\]

The training objective enforces accurate reconstruction at every prefix scale:

\begin{equation}
\label{eq:msae_loss}
\mathcal{L}(x) =
\sum_{m \in \mathcal{M}} \alpha_m
\| x - \hat{x}^{(m)} \|_2^2
+
\lambda \| z \|_1.
\end{equation}

This nested objective creates an ordered latent space and imposes a hierarchical structure: early latents must reconstruct the input independently using limited capacity and therefore capture broad, general factors, while later latents refine the representation with more specialized features. Crucially, later latents cannot change earlier prefixes, preventing feature absorption and preserving hierarchical structure.

Empirical results in large language models show that Matryoshka SAEs significantly reduce feature absorption, lower feature composition and improved interpretability metrics, with only minor trade-offs in reconstruction fidelity compared to standard SAEs, while maintaining competitive downstream performance \cite{Bussmann2025Matryoshka}. In this work, we apply this hierarchical sparse decomposition to recommender system embeddings derived from interaction data, where semantic structure emerges from collaborative behavior rather than linguistic context.

\section{Exploratory Data Analysis}

\subsection{Amazon Fashion Dataset}
We conducted our project on the Amazon Fashion dataset, a large-scale e-commerce dataset containing product metadata and user-item interaction records. The dataset consists of two primary components: an item metadata table and a ratings table.

The item metadata contains 202,642 products, each labeled with between 2 and 8 sub-categories. In the raw dataset, there were 1,051 unique sub-categories. To reduce extreme sparsity and eliminate underrepresented categories, we removed categories associated with fewer than 40 items. This pruning step reduced the number of categories to 500, yielding a more stable structure for downstream analysis.

\begin{figure}[H]
    \centering
    \includegraphics[width=0.7\linewidth, trim=0 0 0 18, clip]{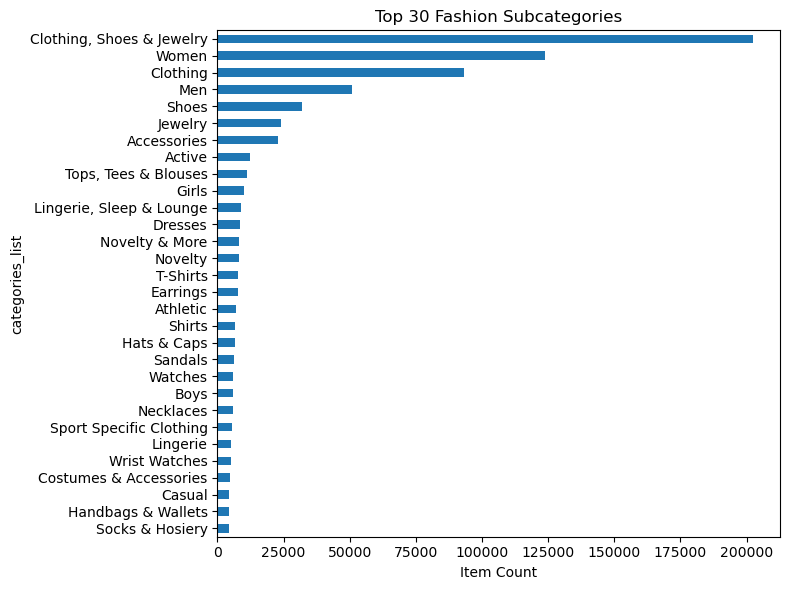}
    \caption{The top 30 fashion sub-categories in the Amazon Fashion dataset.}
    \label{fig:amazon_fahion_sub-categories}
\end{figure}

The ratings dataset contains 9,127,377 interaction records, with 1,096,901 unique users and 198,771 unique items. To ensure sufficient signal for collaborative filtering, we pruned users with fewer than 3 interactions and items with fewer than 5 interactions. After pruning, the dataset contains 1,089,923 users and 182,207 items.

These preprocessing steps balance two competing objectives: retaining large-scale interaction diversity while mitigating extreme sparsity that would otherwise hinder matrix factorization training.

The original dataset provides explicit ratings on a 1–5 scale. In our setting, we treat the data as implicit feedback and binarize interactions, considering any observed rating as a positive interaction. This aligns with standard large-scale collaborative filtering practice in sparse e-commerce settings, where the presence of interaction is typically more informative than the rating magnitude.

\subsection{Interaction Sparsity and Structural Properties}
Before pruning, given 9,127,377 observed interactions among 1,096,901 users and 198,771 items, the resulting matrix density is  $4.19 \times 10^{-5}$ (0.0042\%). This extreme sparsity is characteristic of real-world recommender systems and highlights the challenge of learning meaningful latent representations from limited per-user interaction signals.

After pruning low-activity users and low-frequency items, the final interaction matrix has dimensions of $(1{,}089{,}923 \times 182{,}207)$, with non-zero entries corresponding to observed interactions.

In addition to the user-item matrix, we construct an item-category matrix of size $(500 \times 182{,}207)$, indicating category membership for each item. Because each item belongs to multiple categories, the category structure is inherently hierarchical and overlapping.

This multi-label structure is particularly relevant for interpretability analysis, as the dataset exhibits fine-grained category distinctions. As a result, latent representations may encode both broad concepts (e.g., women's clothing) and highly specific refinements (e.g., floral summer maxi dresses).

\subsection{Sparse Matrix Construction}
Training matrix factorization on a dataset of this scale presents computational challenges due to the size of the interaction matrices. To efficiently handle the large and sparse structure, we construct the interaction matrices using the COO (Coordinate) sparse matrix format in Python.

The COO representation stores only non-zero entries as triplets $(row, column, value)$, dramatically reducing memory usage. This representation is later converted into formats optimized for matrix multiplication during training. The sparse implementation enables scalable MF training without materializing dense matrices of prohibitive size.

This design choice is essential for preserving the full scale of the dataset while maintaining computational feasibility.

\subsection{Implications for Representation Learning}
The Amazon Fashion dataset presents a challenging and realistic testbed for hierarchical interpretability. The presence of multi-level and overlapping category structures suggests that collaborative filtering embeddings may encode both general and specific product concepts.

\begin{figure}[H]
    \centering

    \begin{minipage}{0.48\linewidth}
        \centering
        \includegraphics[width=\linewidth, trim=0 0 0 18, clip]{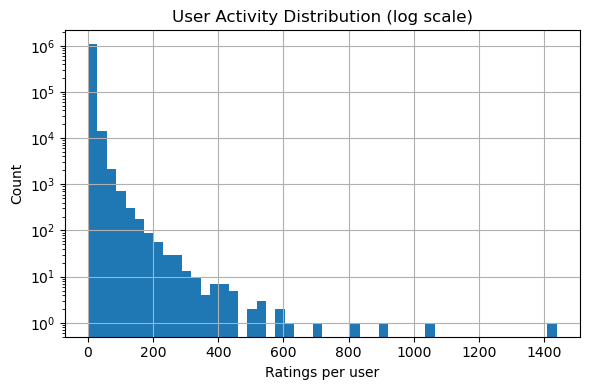}
        \caption{User Activity Distribution (log scale)}
        \label{fig:amazon_user_activity}
    \end{minipage}
    \hfill
    \begin{minipage}{0.48\linewidth}
        \centering
        \includegraphics[width=\linewidth, trim=0 0 0 18, clip]{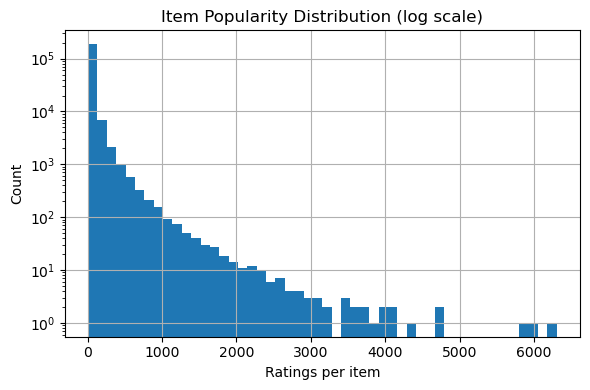}
        \caption{Item Popularity Distribution (log scale)}
        \label{fig:amazon_item_popularity}
    \end{minipage}

\end{figure}

However, the extreme sparsity of interactions and the long-tailed distribution of user activity as seen in Figure~\ref{fig:amazon_user_activity} introduce statistical signals such as popularity effects, which may become entangled with semantic factors in dense embeddings.

This combination of hierarchical metadata structure and interaction-driven statistical effects makes the dataset well-suited for evaluating whether hierarchical sparse autoencoders can disentangle semantically coherent latent factors from entangled collaborative filtering representations.

\section{Methodology}
This section describes the experimental pipeline used to study monosemanticity in recommender system embeddings. Our approach follows a post-hoc interpretability paradigm: we first train a large-scale matrix factorization model to obtain dense user and item embeddings, and subsequently apply a Matryoshka Sparse Autoencoder (MSAE) to analyze their internal structure. Importantly, sparse autoencoder training is fully decoupled from recommendation performance and serves purely as an interpretability tool.

\subsection{Matrix Factorization with Implicit Feedback}
We train a matrix factorization (MF) model on the pruned Amazon Fashion dataset using implicit feedback and on 40\% of the users for computational efficiency. After binarization, each observed interaction is treated as a positive signal, while unobserved user–item pairs serve as potential negatives.  

Matrix factorization represents each user and item with a dense embedding vector of fixed dimension $d =20$, which was chosen arbitrarily. The relevance score between a user and an item is computed as the dot product between their embeddings. Intuitively, users and items with aligned embeddings receive higher predicted interaction scores.

Because the interaction matrix is extremely sparse, we employ negative sampling during training - for each observed interaction, a subset of unobserved user-item pairs is sampled. The model is trained to assign higher scores to observed interactions than to sampled negatives using a binary cross-entropy loss with sigmoid activation, combined with L2 regularization on the user and item embeddings.

After training, we retain the learned item embeddings and treat them as fixed representations for the interpretability analysis described below. The sparse autoencoder is trained independently and does not influence the recommendation model or its training.

\subsection{Matryoshka Sparse Autoencoder}
To analyze the internal structure of the learned item embeddings, we train a Matryoshka Sparse Autoencoder (MSAE) on the fixed item embeddings produced by the matrix factorization model.

A standard sparse autoencoder learns a compressed representation of each embedding while enforcing sparsity, encouraging only a small subset of latent units to activate for any given item. However, when the latent dictionary becomes large, standard SAEs can suffer from feature splitting and absorption, where high-level concepts fragment or are overwritten by more specialized features.

Matryoshka SAEs address this limitation by learning an ordered latent representation. Instead of training a single flat sparse dictionary, the model enforces accurate reconstruction at multiple nested prefix sizes. In practice, this means that the first subset of latent units must already provide a reasonable reconstruction of the input, while progressively larger subsets refine it. The prefixes we chose to train are \{5, 10, 15, 20, 25, 30, 35, 40, 45, 50\}.

This hierarchical training mechanism encourages early latent units to capture broad and general factors, while later units specialize in more fine-grained refinements. Because each prefix must independently reconstruct the embedding, later units cannot simply replace earlier ones, reducing feature absorption and preserving hierarchical structure.

We train the Matryoshka SAE using a reconstruction objective combined with sparsity regularization, as described in Equation~\ref{eq:msae_loss}. Concretely, the training loss combines four terms: a weighted MSE reconstruction loss summed across all prefix sizes, an L1 sparsity penalty on the latent activations, a KL-divergence penalty enforcing a target sparsity level, and an inner product preservation loss that encourages the reconstructed embeddings to maintain the same user--item relevance scores as the original MF embeddings. Importantly, the autoencoder is trained purely on the learned item embeddings and does not use any category labels or metadata during training. Any semantic structure that emerges in the latent space therefore reflects information already encoded in the collaborative filtering representations.

Table~\ref{tab:hp} summarizes the final hyperparameters used for both the MSAE and the plain SAE baseline. Both models were tuned using Optuna-based hyperparameter search to maximize the mean monosemanticity score (MS) across neurons.

\begin{table}[H]
\centering
\caption{Final hyperparameters for MSAE and SAE training after hyperparameter tuning.}
\label{tab:hp}
\begin{tabular}{lcc}
\toprule
\textbf{Hyperparameter} & \textbf{MSAE} & \textbf{SAE} \\
\midrule
Input dimension                     & 20                      & 20                   \\
Hidden dimension                    & 50                      & 50                   \\
Prefix sizes                        & \{5,10,\ldots,50\}      & ---                  \\
Epochs                              & 30                      & 30                   \\
Learning rate                       & $1\times10^{-4}$        & $1\times10^{-4}$     \\
Batch size                          & 256 & 256           \\
\bottomrule
\end{tabular}
\end{table}

\subsection{Monosemanticity Evaluation}
To quantify the semantic coherence of individual latent neurons, we adapt the Monosemanticity Score (MS) introduced by Pach et al. (2025) \cite{Pach}. In the original formulation, MS measures the activation-weighted average pairwise similarity between images that strongly activate a given neuron.

In our setting, items replace images in the original formulation. For each latent neuron, we identify items with high activation values and compute the activation-weighted average pairwise similarity between them. Semantic similarity is computed using cosine similarity between the corresponding item embedding vectors obtained from the matrix factorization model.

Neurons that consistently activate on semantically similar items receive higher monosemanticity scores, while neurons that activate on diverse item sets receive lower scores. This provides a quantitative measure of monosemanticity in recommender embeddings.

\subsection{Latent Concept Labeling}
In addition to quantitative evaluation, we perform qualitative analysis of latent neurons. For each high-activation latent neuron, we extract the top-activating items and provide their metadata to a large language model (LLM) for semantic labeling.

The LLM generates candidate human-interpretable concept descriptions based on shared attributes among the activated items. This process enables us to associate latent neurons with descriptive semantic labels, facilitating interpretability and supporting the quantitative monosemanticity analysis. The LLM prompts can be seen in appendix A.

\section{Results \& Analysis}

\subsection{Comparison between SAE \& MSAE}

To assess whether the hierarchical structure imposed by the Matryoshka training objective comes at a cost to recommendation quality, we trained a plain SAE with a matching hidden dimension of 50 and evaluated both models on the same held-out validation set. Because the autoencoder operates purely as a post-hoc representation analysis tool, recommendation scores are computed using the reconstructed item embeddings produced by each model while keeping the original matrix factorization user embeddings fixed. The reported metrics therefore measure how well each reconstruction preserves the ranking behavior of the underlying MF recommender rather than evaluating a new recommendation model.

Table~\ref{tab:sae_msae} summarizes the results. The MSAE achieves a marginally higher AUC of 0.7754 compared to 0.7685 for the SAE. Ranking metrics tell a similar story: MSAE yields a Precision@10 of 0.0012, Recall@10 of 0.0059, and NDCG@10 of 0.0033, versus 0.0011, 0.0057, and 0.0032 for the SAE respectively. The absolute values of all ranking metrics are near zero, which is expected given the extreme sparsity of the Amazon Fashion dataset (density $\approx 0.004\%$) and the small number of held-out positives per user.

\begin{table}[H]
\centering
\caption{SAE vs.\ MSAE recommendation metrics on Amazon Fashion.}
\label{tab:sae_msae}
\begin{tabular}{lcc}
\toprule
\textbf{Metric} & \textbf{SAE} & \textbf{MSAE} \\
\midrule
AUC             & 0.7685 & 0.7754 \\
Log Loss        & 0.5918 & 0.5910 \\
Accuracy        & 0.7121 & 0.7209 \\
Precision@10    & 0.0011 & 0.0012 \\
Recall@10       & 0.0057 & 0.0059 \\
NDCG@10         & 0.0032 & 0.0033 \\
\bottomrule
\end{tabular}
\end{table}

\begin{figure}[H]
    \centering
    \includegraphics[width=\linewidth]{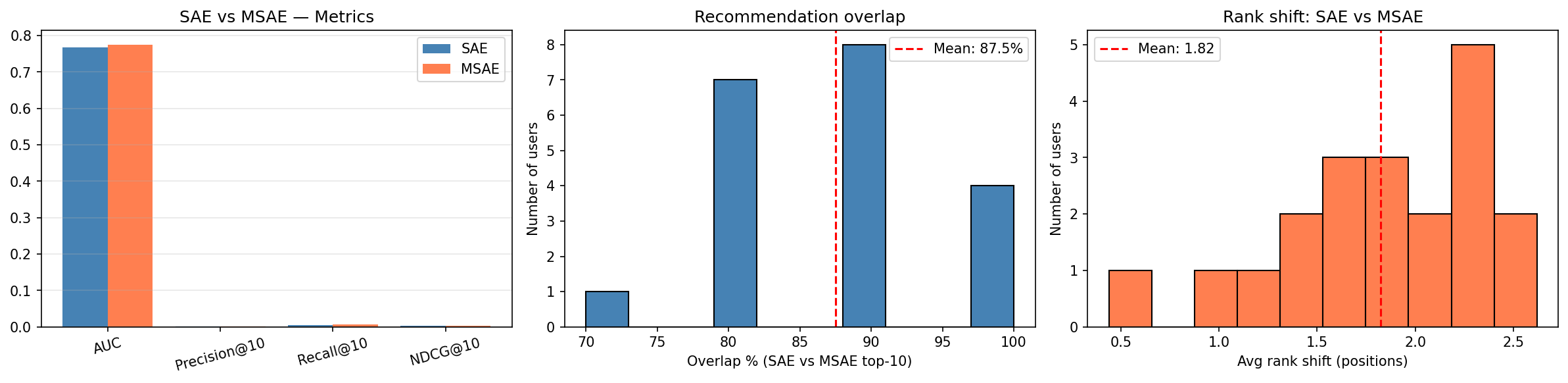}
    \caption{SAE vs.\ MSAE comparison: recommendation metrics (left), top-10 list overlap distribution (centre), and rank shift distribution (right) across 20 sampled users.}
    \label{fig:sae_msae_comparison}
\end{figure}

Beyond aggregate metrics, we examined the overlap between the top-10 recommendation lists produced by each model across 20 sampled users. The mean list overlap was 87.5\%, and the mean absolute rank shift among shared items was 1.82 positions, as shown in Figure~\ref{fig:sae_msae_comparison}. These results indicate that the Matryoshka objective does not strongly alter the ranking behavior of the underlying MF embeddings: the two models agree on the vast majority of recommended items and differ only marginally on their relative ordering. Taken together, the evidence suggests that the hierarchical constraint introduced by MSAE incurs no measurable cost to recommendation fidelity, while providing the additional benefit of a nested latent structure amenable to monosemanticity analysis at multiple granularities.

\subsection{Qualitative Results}

\subsubsection{Monosemanticity Score Distribution of Latent Neurons}

To quantify semantic specialization across the learned latent space, we computed the Monosemanticity Score (MS) for each of the 50 MSAE neurons using a subsample of 5{,}000 items. MS scores range from approximately 0.075 to 0.201, indicating a meaningful spread of semantic coherence across the latent dictionary.

\begin{figure}[H]
\centering
\includegraphics[width=0.55\linewidth, trim=0 0 0 20, clip]{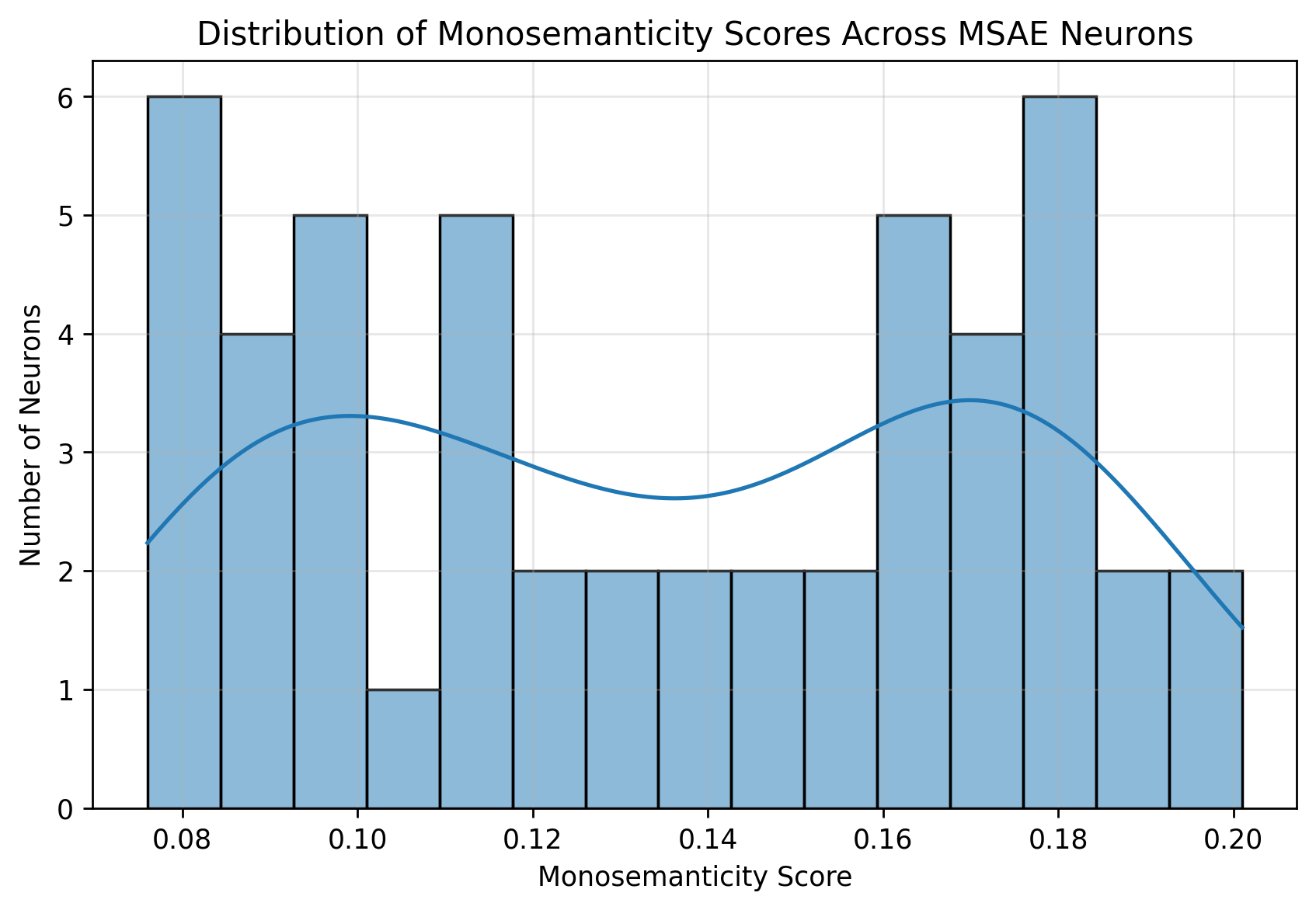}
\caption{
Distribution of Monosemanticity Scores across latent neurons.
Higher scores indicate stronger specialization of neurons toward a coherent semantic concept.
The observed distribution suggests that several neurons capture moderately specialized semantic features.
}
\label{msae_ms_score_distribution_kde}
\end{figure}

Figure~\ref{msae_ms_score_distribution_kde} shows the distribution of MS scores across neurons. The distribution is notably bimodal, with one cluster concentrated in the range 0.08--0.11 and a second cluster at 0.16--0.19. This pattern suggests the presence of two qualitatively distinct neuron populations: a majority of broadly tuned neurons that activate across diverse item types, and a smaller group of more specialized neurons that respond consistently to semantically coherent item subsets.

The top-scoring neurons are N41 (0.201), N44 (0.196), N43 (0.192), N40 (0.188), and N36 (0.183), while the lowest-scoring neurons are N3 (0.076), N1 (0.077), and N0 (0.079). Notably, the top-scoring neurons are predominantly drawn from the higher-indexed portion of the latent dictionary (neurons 36--49), while the lowest-scoring neurons are concentrated among the earliest indices (0--3). This pattern is consistent with the Matryoshka training objective: early prefix neurons must reconstruct the full embedding with limited capacity and therefore capture broad variance, while later neurons are free to specialize.

It is important to note that the MS score measures activation-weighted pairwise item similarity in embedding space, and therefore reflects the geometric coherence of a neuron's activation set rather than the interpretability of its semantic label. A neuron can receive a consistent and interpretable LLM-generated label while still scoring low on MS if the items it activates on are spread across a broad region of the embedding space. This is the case for N3, which receives a women's apparel label from the labeling pipeline yet ranks among the lowest MS scores: its top-activating items share a coherent category label but are geometrically dispersed in the MF embedding space. MS score and label coherence are therefore complementary but distinct notions of monosemanticity.

\begin{figure}[H]
\centering
\includegraphics[width=0.8\linewidth, trim=0 0 0 20, clip]{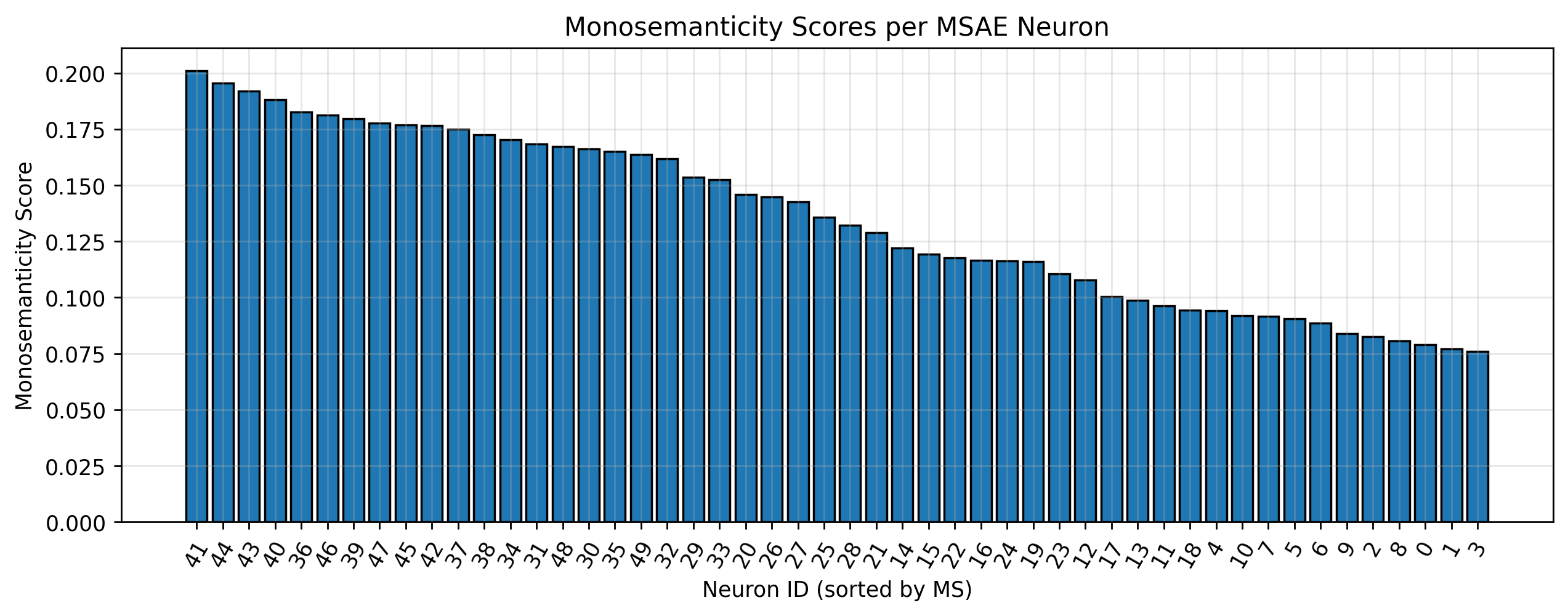}
\caption{
Monosemanticity Scores per neuron sorted by magnitude.
Neurons with higher scores exhibit stronger semantic coherence across their top activating items, supporting their interpretability and suitability for causal intervention analysis.
}
\label{msae_ms_score_per_neuron_sorted}
\end{figure}

Figure~\ref{msae_ms_score_per_neuron_sorted} presents the MS scores sorted by neuron index. Several neurons exhibit relatively higher scores, suggesting that they respond consistently to specific semantic categories.

Most of the neurons selected for intervention in the following subsection fall within the higher end of the MS distribution, supporting the interpretation that they capture coherent semantic concepts related to women's fashion categories.

\subsubsection{Semantic Gradient Across the Latent Dictionary}

Qualitative inspection of the LLM-generated neuron labels reveals a semantic gradient that mirrors the quantitative MS score pattern. Neurons 0--21 receive broad labels such as ``Clothing, Shoes \& Jewelry'' with no consistent subcategory specialization, reflecting their role as general-purpose reconstructors in the early prefixes. Neurons 22--35 show mild demographic focus, with labels suggesting partial alignment with gendered product categories. Neurons 36--49 exhibit the most differentiated labels, including specific references to women's apparel subcategories, footwear types, and accessory segments.

Among the higher-indexed neurons, several exhibit strong and consistent associations with women-oriented product categories. Neurons N3, N32, N39, N41, N45, and N48 consistently activate on women's apparel, lingerie, dresses, and footwear items across their top-activating sets, forming the basis for the gender-sensitive axis analyzed in the following subsection. The presence of this axis exclusively among higher-indexed neurons further supports the interpretation that the Matryoshka hierarchy encourages progressive semantic refinement rather than flat, undifferentiated feature allocation.

These findings suggest that collaborative filtering embeddings trained on implicit interaction data contain recoverable hierarchical semantic structure, and that the Matryoshka training objective provides a principled mechanism for exposing it.

\subsection{Latent Gender Axis Discovery in the Recommendation Representation}
\subsubsection{Discovery of Gender-Associated Latent Neurons}
To investigate semantic structure in the learned latent representation, we applied the automatic neuron labeling pipeline described in Appendix~A. Using the top-activating items of each neuron, Claude Opus 4.5 inferred broad semantic labels grounded in repeated category patterns.
Among the 50 latent neurons analyzed, several neurons exhibited consistent associations with women-oriented product categories such as women's apparel, dresses, lingerie, and footwear. Based on these labels, we selected a subset of neurons strongly associated with women dominant categories:

\[
\mathcal{N}_{gender} = \{3, 32, 39, 41, 45, 48\}.
\]

This subset corresponds to 6 out of the 50 latent neurons learned by the autoencoder, representing approximately 12\% of the latent feature space.
Inspection of the top activating items confirmed that these neurons consistently respond to products belonging to women's apparel and accessories categories, suggesting that they encode a latent semantic direction related to gender-oriented fashion preferences.
This observation suggests the presence of a \textbf{gender sensitive axis} within the recommender system's representation space.

\subsubsection{Causal Intervention on Gender Neurons}

To test whether the identified neurons causally influence recommendation behavior, we performed controlled interventions by modifying their activations during inference.

For each user representation $z$, we applied scaling to the selected neurons:

\[
z_i' =
\begin{cases}
\alpha z_i & i \in \mathcal{N}_{gender} \\
z_i & \text{otherwise}
\end{cases}
\]

where $\alpha$ controls the intervention strength.

Two intervention conditions were evaluated:

\begin{itemize}
    \item \textbf{Amplify}: $\alpha = 1.5$
    \item \textbf{Suppress}: $\alpha = 0.5$
\end{itemize}

Experiments were conducted on a randomly sampled set of 50 users (random seed = 42), for computational simplicity. For each user, we computed Top-K recommendations with $K = 50$.
Recommendations were evaluated under three conditions:

\begin{itemize}
    \item Baseline (no intervention)
    \item Amplified gender neurons
    \item Suppressed gender neurons
\end{itemize}

This setup allows us to measure how manipulating the activation of gender associated neurons affects recommendation outcomes.

\subsubsection{Impact on Recommendation Composition}
\begin{figure}[H]
\centering
\includegraphics[width=0.70\linewidth, trim=0 0 0 19, clip]{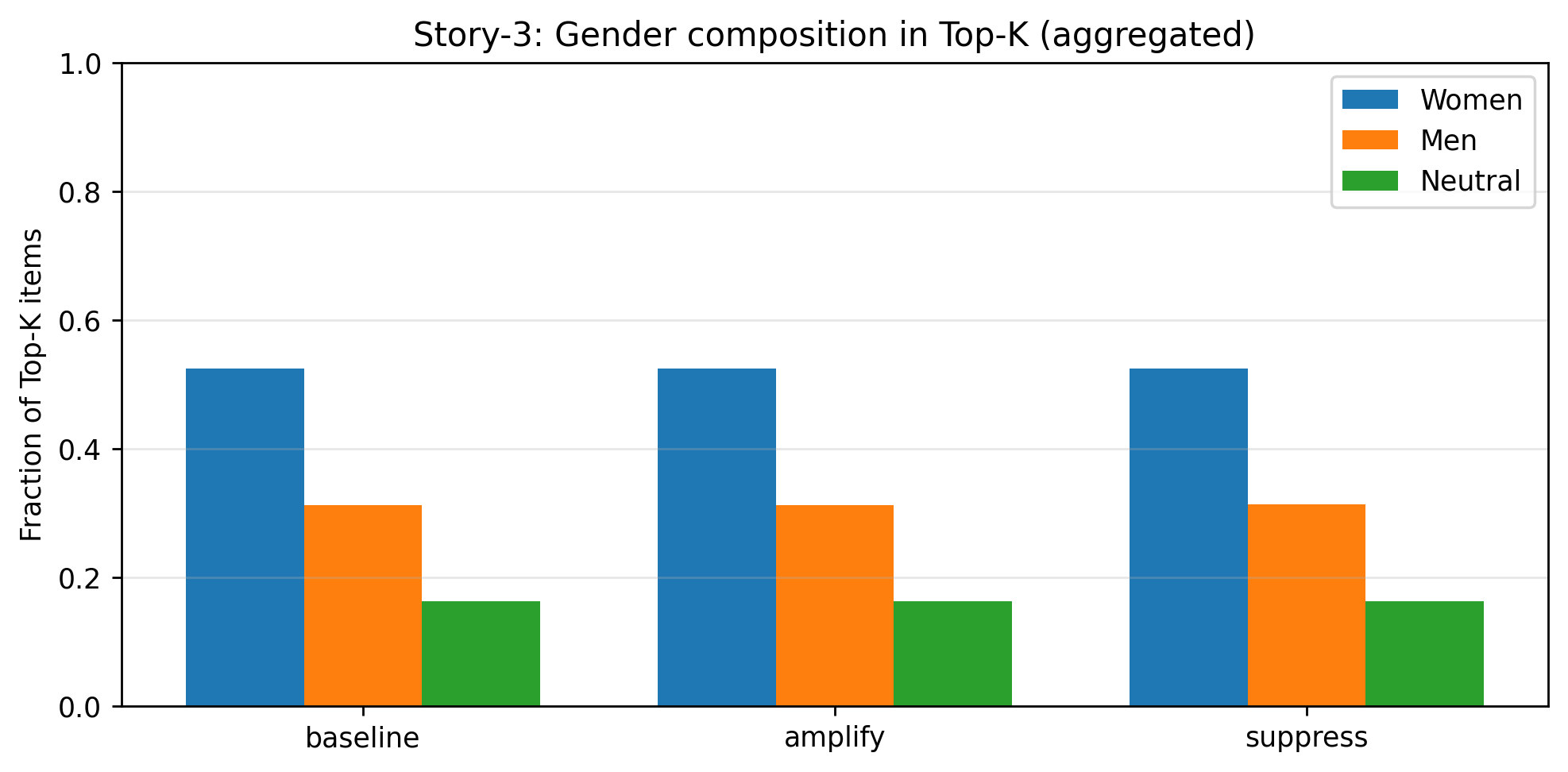}
\caption{
Gender composition of Top-$K$ recommendations under different intervention conditions.
The plot shows the fraction of recommended items associated with women, men, and neutral categories for baseline recommendations and neuron interventions.
Amplifying gender-associated neurons slightly increases the share of women-oriented products, while suppressing them produces the opposite effect.
}
\label{fig_gender_composition_topk}
\end{figure}

Figure~\ref{fig_gender_composition_topk} shows the aggregated gender composition of Top-K recommendations across the evaluated users.

Women-associated products constitute the majority of recommendations across all conditions, accounting for approximately 52--54\% of recommended items, while men-associated products account for roughly 30--32\%.

Despite relatively small shifts in aggregate percentages, intervention effects are observable at the item level. Amplification slightly increases the presence of women associated items for some users, while suppression increases the proportion of neutral or men associated items in others.

\begin{figure}[H]
\centering
\includegraphics[width=0.65\linewidth, trim=0 0 0 20, clip]{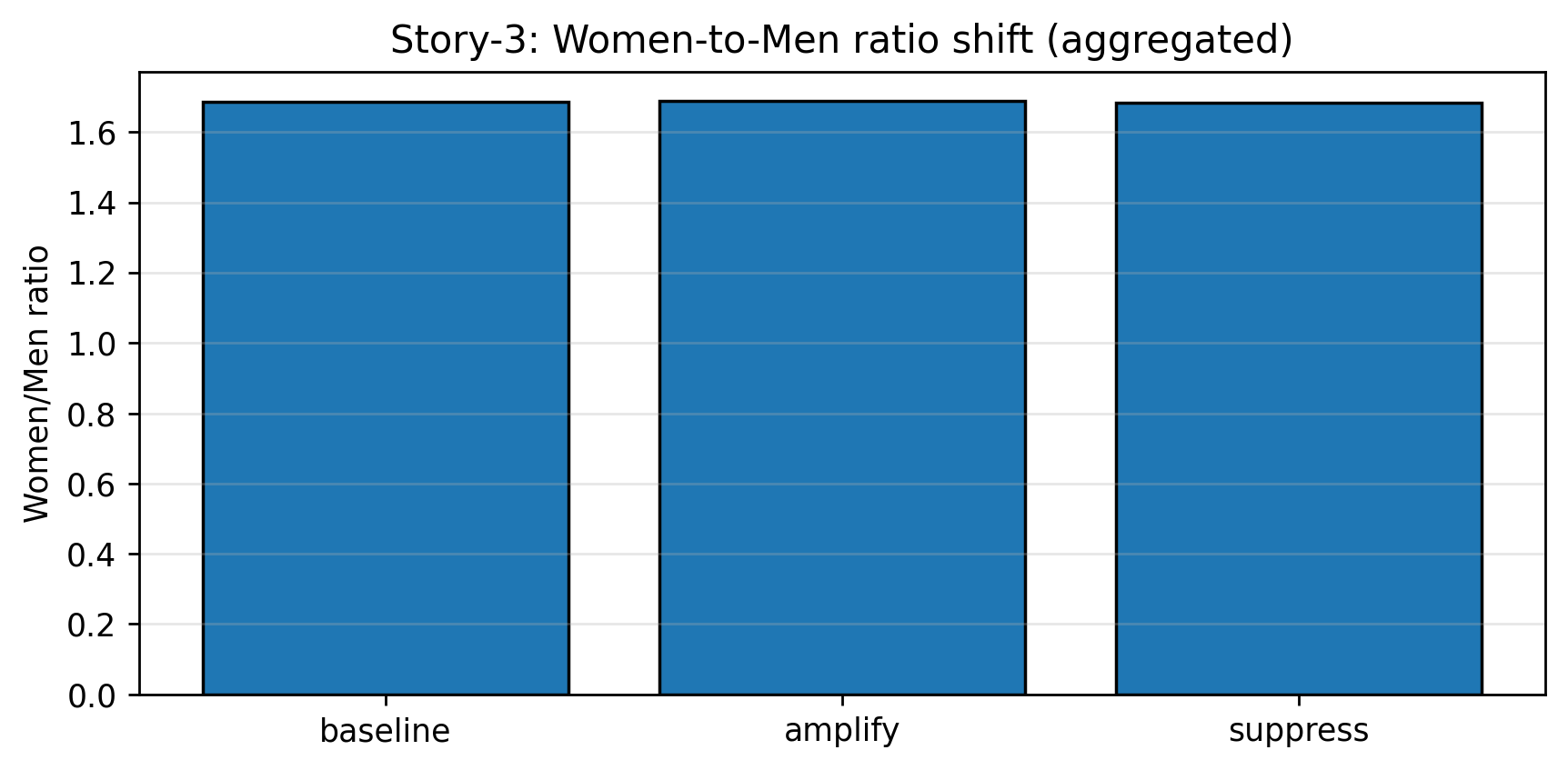}
\caption{
Women to men recommendation ratio under baseline and intervention conditions.
Although neuron scaling slightly shifts the gender composition of recommendations, the overall ratio remains relatively stable, indicating that the intervention modifies recommendation preferences without dominating the entire ranking process.
}
\label{fig_gender_ratio_shift}
\end{figure}

The ratio of women to men recommendations remains relatively stable across conditions (Figure~\ref{fig_gender_ratio_shift}), indicating that the discovered neurons influence recommendation composition but do not dominate the entire recommendation process.

Importantly, recommendation lists remain largely stable across interventions. The average overlap between baseline and intervention recommendations exceeds 0.96, indicating that neuron scaling modifies rankings rather than replacing the entire recommendation set.

\begin{figure}[H]
\centering
\includegraphics[width=0.8\linewidth, trim=0 0 0 20, clip]{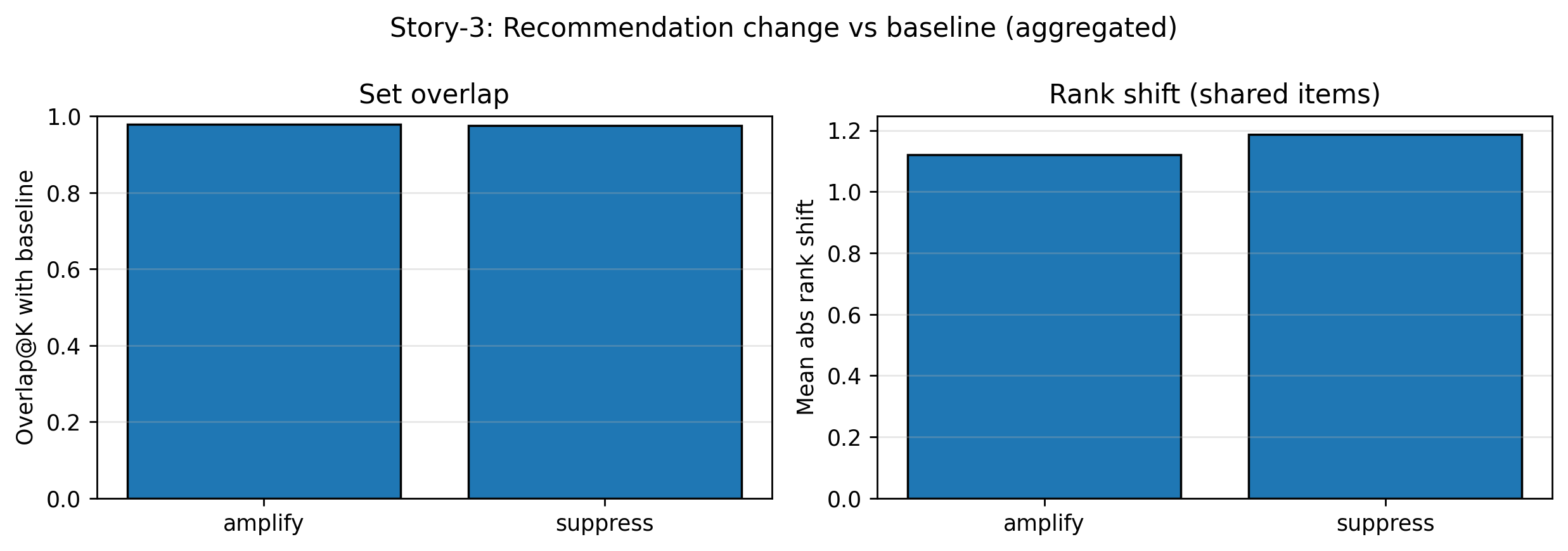}
\caption{
Recommendation stability under neuron intervention.
Left: Overlap@50 between baseline and intervention recommendation lists.
Right: Average absolute rank shift among shared items.
High overlap values ($>0.96$) indicate that neuron interventions primarily adjust item ranking rather than replacing the recommendation set.
}
\label{fig_recommendation_shift}
\end{figure}

Figure~\ref{fig_recommendation_shift} further illustrates that the average absolute rank shift among shared items is modest (approximately 1 rank position), demonstrating that neuron interventions primarily adjust relative ranking rather than introducing entirely new items.

The aggregate behavioral metrics for each intervention condition are summarized in Table~\ref{tab:intervention_metrics}.

\begin{table}[H]
\centering
\small
\begin{tabular}{lcccccc}
\hline
\textbf{Condition} & \textbf{Women \%} & \textbf{Men \%} & \textbf{Overlap@50} & \textbf{Rank Shift} & \textbf{KL} & \textbf{Entropy} \\
\hline
Baseline  & 0.53 & 0.31 & 1.00 & 0.00 & 0.00 & 3.30 \\
Amplified & 0.54 & 0.30 & 0.97 & 1.10 & 0.067 & 3.31 \\
Suppressed & 0.52 & 0.32 & 0.97 & 1.20 & 0.074 & 3.30 \\
\hline
\end{tabular}
\caption{Aggregate recommendation metrics under neuron intervention.}
\label{tab:intervention_metrics}
\end{table}

\subsubsection{Category Distribution Shifts}

To better understand how the intervention modifies recommendation behavior, we analyzed the distribution of product categories within the Top-K recommendations.

\begin{figure}[H]
\centering
\includegraphics[width=0.8\linewidth, trim=0 0 0 20, clip]{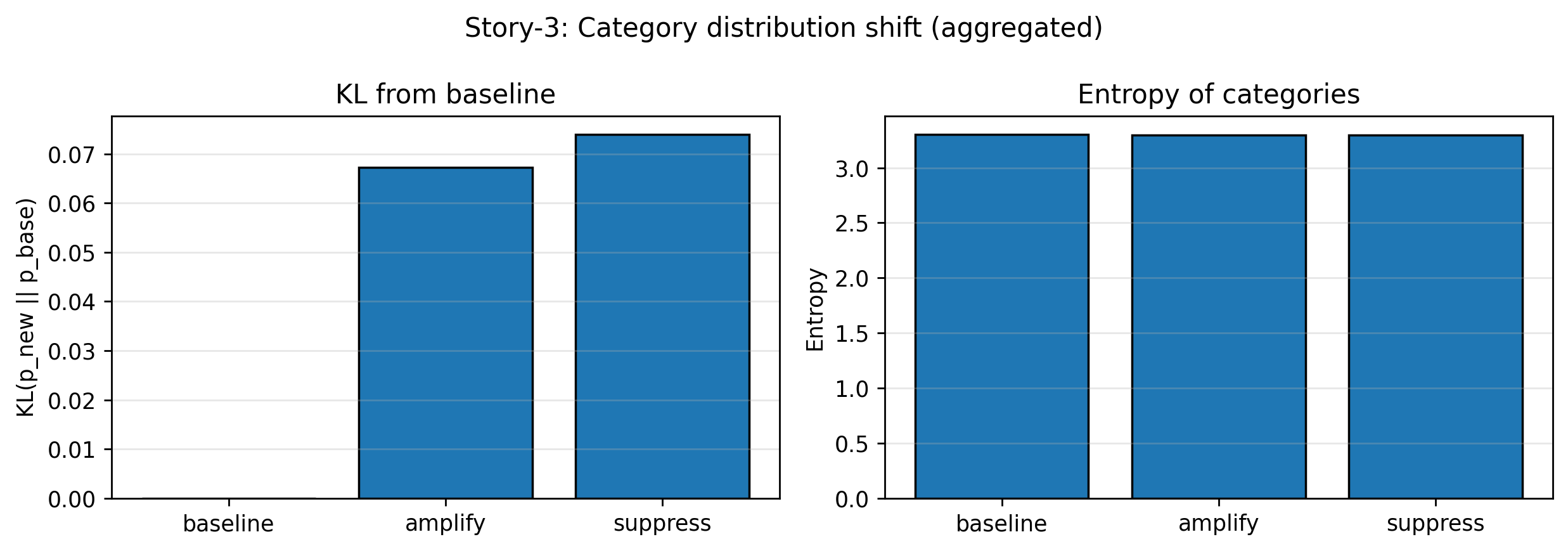}
\caption{
Category distribution shift induced by neuron intervention.
The left panel shows KL divergence between baseline and intervention category distributions, while the right panel reports entropy of the category distribution.
Both interventions introduce moderate distribution shifts while maintaining similar overall diversity.
}
\label{fig_category_distribution_shift}
\end{figure}

Figure~\ref{fig_category_distribution_shift} shows the KL-divergence between category distributions under intervention and baseline conditions. Both amplification and suppression introduce moderate distribution shifts, with KL divergence values around 0.07.

Despite these shifts, overall category diversity remains stable, as indicated by nearly identical entropy values across conditions.

\begin{figure}[H]
\centering
\includegraphics[width=0.8\linewidth, trim=0 0 0 20, clip]{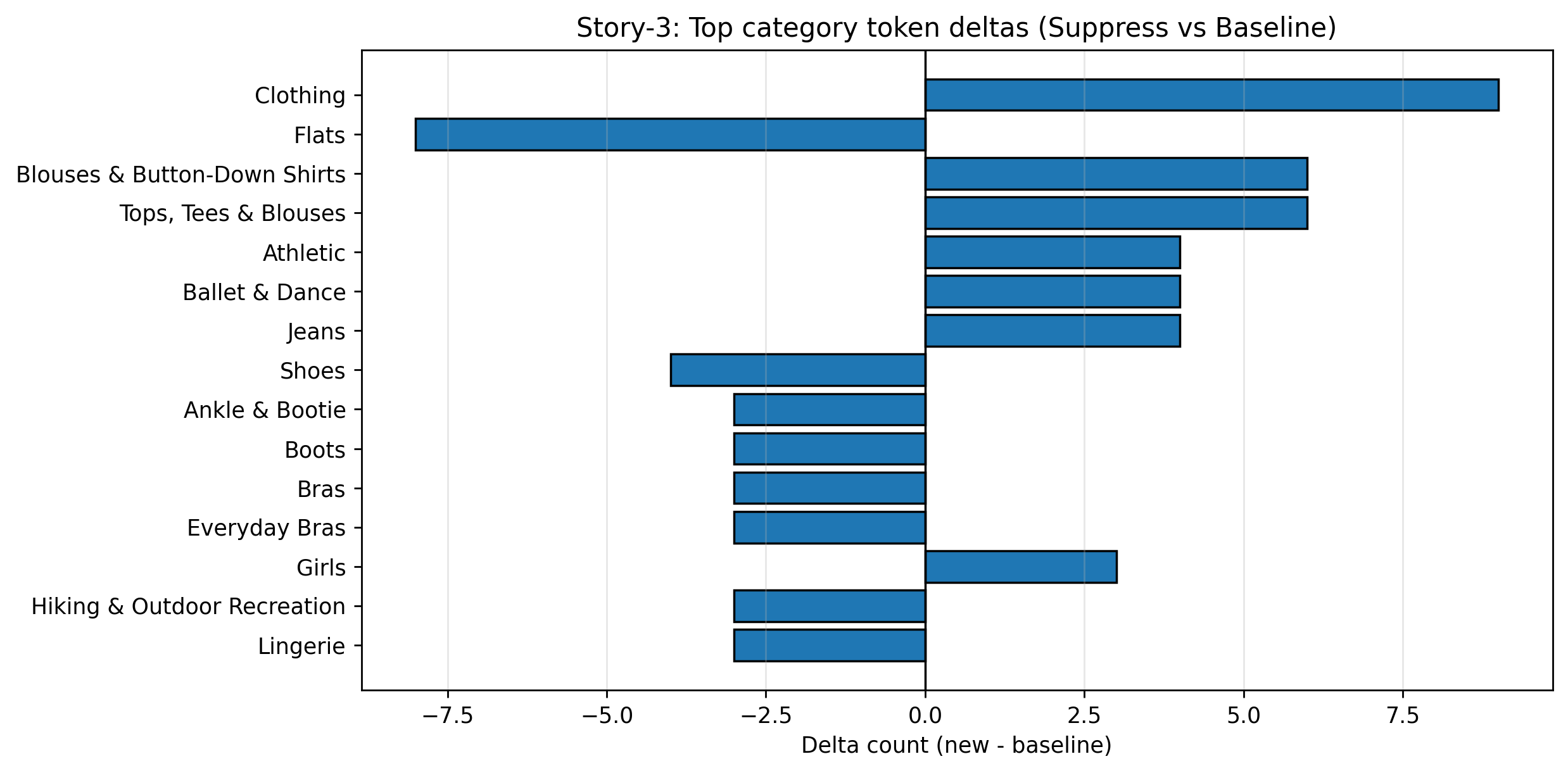}
\caption{
Largest category token frequency changes relative to the baseline recommendation distribution.
Positive values indicate categories that become more frequent after intervention, while negative values indicate categories that decrease.
Suppressing gender associated neurons reduces strongly women-oriented fashion categories such as lingerie and bras, while increasing more neutral apparel categories.
}
\label{fig_category_token_deltas}
\end{figure}

A closer inspection of category token changes (Figure~\ref{fig_category_token_deltas}) reveals interpretable behavioral patterns. Suppressing gender neurons increases the frequency of more neutral categories such as \textit{Clothing}, \textit{Athletic}, and \textit{Jeans}, while decreasing categories strongly associated with women's fashion, including \textit{Bras}, \textit{Lingerie}, and \textit{Flats}.

These results suggest that the identified neurons encode preferences for specific women's apparel categories rather than simply reflecting a global gender label.

\subsubsection{Intervention Response Curve}
To further evaluate the causal role of these neurons, we examined the effect of progressively scaling their activations.

\begin{figure}[H]
\centering
\includegraphics[width=0.55\linewidth, trim=0 0 0 20, clip]{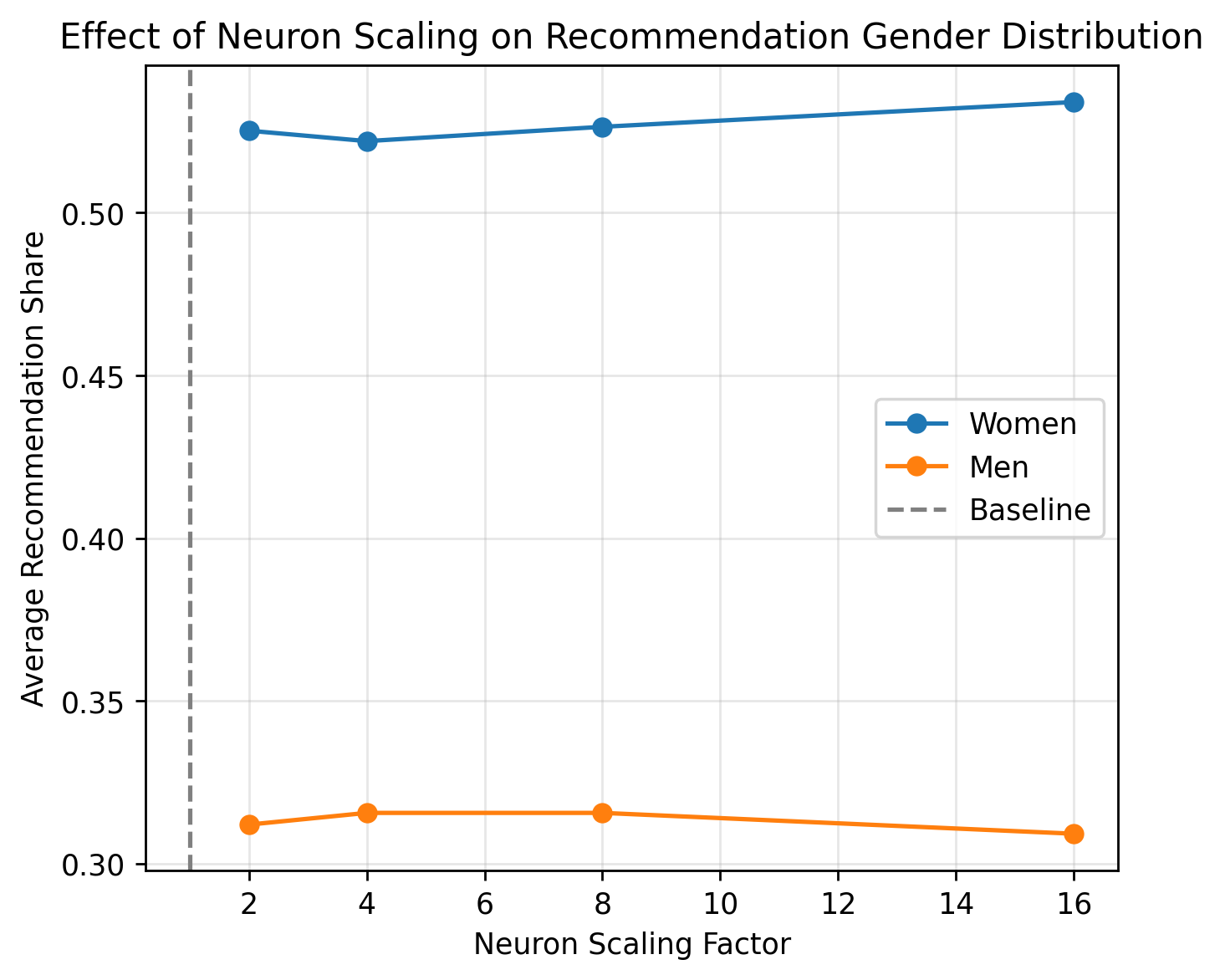}
\caption{
Effect of progressively scaling gender associated neurons on recommendation composition.
Increasing the scaling factor gradually increases the share of women-oriented recommendations while slightly decreasing the share of men oriented items.
The consistent trend across scaling factors suggests that the identified neurons form a controllable semantic axis within the latent representation.
}
\label{intervention_response_curve}
\end{figure}

Figure~\ref{intervention_response_curve} shows the average gender composition of recommendations under different scaling factors.

Increasing the scaling factor gradually increases the share of women associated recommendations, while slightly decreasing the share of men associated items. The effect remains stable across a range of scaling factors (2–16), indicating that the intervention produces a consistent directional shift in recommendation outcomes.

This behavior supports the hypothesis that the selected neurons form part of a controllable latent axis associated with gender related product categories.

\subsubsection{Interpretability Findings}
Taken together, the experiments provide consistent evidence that the MSAE learns semantically meaningful and causally relevant latent features within the recommender system.

First, the automatic labeling pipeline revealed that several neurons consistently activate for women's apparel categories such as dresses, lingerie, footwear, and accessories. This semantic pattern suggests the presence of a latent gender sensitive direction in the representation space.

Second, causal interventions on these neurons demonstrate that modifying their activations systematically alters recommendation outcomes. Amplifying the selected neurons increases the relative presence of women associated products in recommendation lists, while suppressing them produces the opposite effect. Importantly, these changes occur without dramatically altering the recommendation set, as evidenced by the high overlap with baseline recommendations and small average rank shifts.

Third, analysis of category token frequencies shows that interventions modify recommendations in a semantically interpretable manner. Suppressing gender associated neurons increases the frequency of neutral apparel categories while decreasing categories strongly associated with women's fashion. This provides additional evidence that the neurons capture coherent product semantics rather than arbitrary statistical correlations.

Finally, the intervention response curve demonstrates that the effect is stable across a wide range of scaling factors, confirming that the identified neurons form part of a controllable semantic axis within the latent representation.

Overall, these findings suggest that MSAEs can expose interpretable latent structure in recommender systems, enabling targeted analysis of how specific semantic features influence recommendation behavior. Moreover, these results highlight the potential of sparse autoencoders as a tool for interpretability and controllable recommendation systems.

\section{Discussion}
The results presented in this work provide empirical support for the hypothesis that collaborative filtering embeddings encode recoverable hierarchical semantic structure, and that MSAEs offer a principled mechanism for exposing it. By applying MSAEs to matrix factorization item representations, we have demonstrated that monosemanticity can be extracted from interaction-driven models without the use of metadata during the training phase.

A significant highlight of our analysis is the discovery of a controllable gender-sensitive axis within the Amazon Fashion representation space. The six identified neurons (N3, N32, N39, N41, N45, N48) collectively capture associations with women's apparel subcategories including dresses, lingerie, footwear, and accessories. This allows for a level of post-hoc behavioral control previously unavailable in opaque latent factor models. Unlike standard SAEs, which often suffer from feature splitting or feature absorption, the MSAE's nested objective ensures that early latent units capture broad categories while later units refine these into specialized features.

Furthermore, our causal intervention experiments confirm that these extracted features are not mere statistical artifacts but are functionally grounded. Modifying the activation of gender-associated neurons produced predictable, directional shifts in recommendation composition while maintaining an average overlap greater than 0.96 with the baseline recommendation list. This suggests that the MSAE provides a mechanism for principled intervention that adjusts relative ranking rather than disrupting the underlying collaborative signal. Moreover, the intervention response curve remains stable across a wide range of scaling factors, supporting the interpretation that these neurons form a coherent and controllable semantic axis rather than a set of loosely correlated features.

Our analysis also highlights an important distinction between two complementary notions of monosemanticity. The monosemanticity score, adapted from Pach et al.\cite{Pach}, measures geometric coherence in the embedding space by evaluating how tightly the items activate a given neuron cluster within the underlying MF representation. In contrast, the LLM-generated semantic labels capture the interpretability of a neuron's concept as perceived by an external observer. These two perspectives do not always align. For example, neuron N3 receives a consistent women's apparel label yet scores low on MS, because the items activating the neuron are geometrically dispersed in the embedding space. This dissociation suggests that MS and label coherence capture different and complementary aspects of interpretability, and that a complete picture of monosemanticity in recommender systems may require both.

The comparison between standard SAE and MSAE further supports the practical benefits of hierarchical sparse training. The MSAE achieves marginally superior performance on all reported metrics and the mean list overlap of 87.5\% between the two models' top-10 recommendations, combined with a mean absolute rank shift of only 1.82 positions, confirms that the hierarchical constraint does not meaningfully distort the underlying recommendation behavior. 

Taken together, these results suggest that the opacity of collaborative filtering embeddings is not fundamental: meaningful semantic structure can be recovered post-hoc through hierarchical sparse decomposition. This has practical implications for transparency and auditing in real-world recommender systems, where understanding and controlling the semantic axes encoded in latent representations may be important for fairness, explainability, and principled behavioral intervention.

\section{Limitations \& Future Work}
Despite the encouraging results, several limitations of the current study warrant discussion and motivate directions for future investigation.

\textbf{Scale of the latent dictionary:} The MSAE in this work was trained with a latent dimension of 50, using prefix sizes drawn from \{5, 10, 15, 20, 25, 30, 35, 40, 45, 50\}. While this configuration is sufficient to reveal broad semantic structure and identify axes such as the gender axis observed in our experiments, a larger dictionary may support finer-grained feature decomposition. However, scaling dictionary size introduces well-known challenges including increased computational cost and a heightened risk of feature splitting and absorption. Although the Matryoshka training objective is designed to mitigate these pathologies, whether this property holds at substantially larger scales in recommender system embeddings remains an open empirical question.

\textbf{Single dataset and domain:} All experiments were conducted on the Amazon Fashion dataset, a single domain with particular structural properties including strong gender-based category organization. It is unclear to what extent the observed hierarchical structure and gender axis are domain-specific artifacts, or whether analogous axes emerge in other product domains such as electronics, home goods, or media. Evaluating the approach across multiple datasets would strengthen the generalizability of the findings.

\textbf{Monosemanticity metric:} The MS score provides a useful quantitative measure of geometric coherence, but its relationship to human-perceived interpretability is indirect. As demonstrated by the case of N3, a neuron can receive a coherent LLM-generated label while scoring low on MS. Developing richer evaluation protocols that combine geometric, statistical, and human judgment measures would improve the robustness of monosemanticity assessment in recommender systems.

\textbf{Automated semantic labeling:} The interpretability analysis in this study relies on automated semantic labeling using Claude Opus 4.5. While this approach enables scalable qualitative analysis of latent neurons, it introduces potential biases stemming from the LLM’s own internal representations and training data. Human annotation or hybrid evaluation frameworks combining automated and human validation could provide a more reliable assessment of semantic coherence. 

\textbf{Engineering and scalability considerations:} The computational challenge of sparse matrix construction for datasets of this magnitude is significant. While the COO format mitigates memory usage, training large-scale sparse autoencoders on embeddings derived from billions of interactions and millions of items remains an open engineering hurdle in regards to computational time efficiency.

Addressing these limitations offers several promising directions for future work, including scaling MSAEs to larger latent dictionaries, evaluating the approach across diverse recommendation domains, improving the measurement of monosemanticity, and exploring more scalable training methods for large-scale recommender representations.

\section{Summary \& Conclusions}
In this work, we have bridged the gap between mechanistic interpretability and recommender systems by implementing a hierarchical decomposition framework for latent embeddings. Our application of Matryoshka Sparse Autoencoders to the Amazon Fashion dataset demonstrates that hierarchical sparse representations can successfully expose the layered semantic structure of interaction data.

We adapted the monosemanticity score to quantify feature coherence and implemented an automated labeling pipeline to interpret the resulting latent space, discovering actionable concepts like gender-oriented fashion. Our results confirm that post-hoc interventions on these monosemantic neurons can steer recommendation behavior in a controllable and interpretable manner without retraining the base model. Ultimately, this work provides a foundation for more transparent, explainable, and accountable personalization, enabling developers to audit and refine the internal representations that drive user discovery in modern digital platforms.

\newpage
\appendix
\section{Automatic Neuron Labeling with Claude Opus 4.5}
To scale qualitative interpretability analysis, we employed an automatic neuron-labeling pipeline using Claude Opus 4.5. For each latent neuron, we extracted 25 top-activating items and provided minimal metadata - item name and category path, to the model. The LLM was instructed to infer whether a coherent semantic theme was present and, when applicable, to generate a concise descriptive label (e.g., Clothing, Shoes, Jewelry). If a broad label resembled a previous label then the model added a refined specific subtype label. This procedure enables systematic high-level semantic labeling without manual annotation. The model was prompted using a dual instruction format consisting of a system prompt and a content prompt

The system prompt template was as follows:
\begin{lstlisting}[caption={Claude Opus 4.5 system prompt}]
"""
You are given a list of items that are most strongly associated with a latent neuron from a recommender system. Each item includes a title and includes minimal tags such as a short category path.

Your task is to identify the MOST consistent taxonomy pattern in the list: Provide a BROAD label (high-level theme), e.g. "Women's shoes", "Men's clothing", "Jewelry", "Kids apparel", "Accessories".

Only return "No clear concept" if you cannot find a stable broad theme at all.

Rules:
- broad_label: short high-level theme.
- refined_label: REQUIRED. If broad_label resembles a previous label, refined_label MUST add a specific subtype using category/title patterns (e.g., Women's apparel" -> "Women's apparel - dresses/skirts").
- Avoid exact repeats of broad_label if possible; if unavoidable, keep broad_label and make refined_label different.
- Use ONLY the evidence provided (titles/categories).
- Do NOT hallucinate unseen details.
- Output JSON only.
"""
\end{lstlisting}

The content prompt template was as follows:

\begin{lstlisting}[caption={Claude Opus 4.5 content prompt}]
"""
Neuron ID: {payload['neuron_id']}

Previously used broad labels (avoid exact repeats; if similar, refine with a subtype):
{used_block}

Top-activating items (highest activation first):
<<<
{items_block}
>>>

Return ONLY valid JSON:
{{
  "neuron_id": {payload['neuron_id']},
  "broad_label": "BROAD_LABEL or No clear concept",
  "refined_label": "Refine the broad label with a subtype to avoid repetition",
  "description": "1-2 sentences grounded in repeated title/category patterns",
  "confidence": "low|medium|high",
  "evidence": ["3-6 short repeated patterns you relied on"]
}}
"""
\end{lstlisting}

The resulting labels were used for neuron naming and qualitative analysis throughout the project.

\newpage
\section{Neuron Activation Statistics Before and After Semantic Labeling}

\begin{figure}[h]
\centering
\includegraphics[width=\linewidth, trim=0 0 0 27, clip]{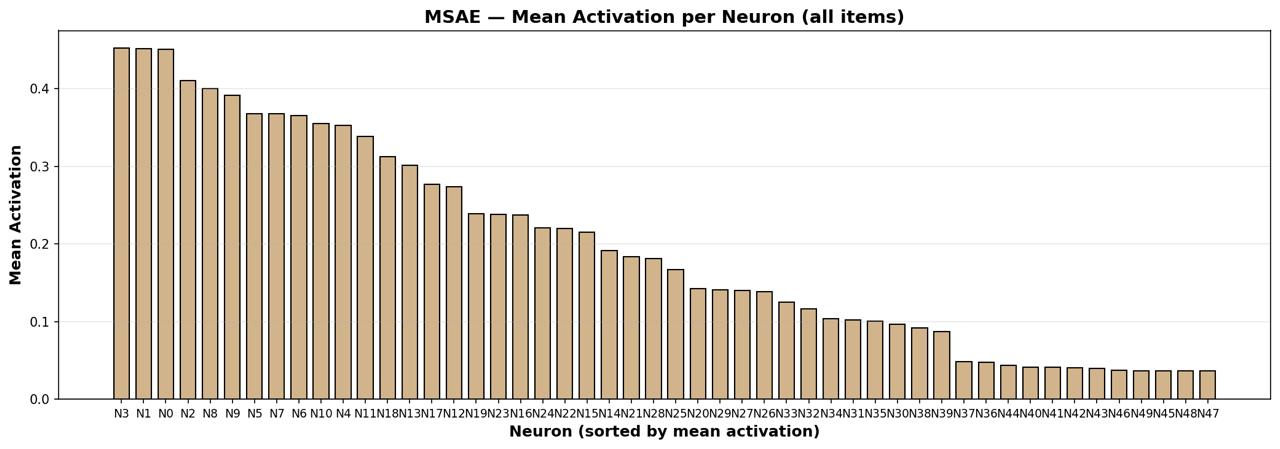}
\caption{
Mean activation magnitude of each latent neuron across all items before semantic labeling. Neurons are sorted by average activation value. This plot illustrates the utilization distribution of the latent dimensions learned by the sparse autoencoder.
}
\label{fig:msae_overall_neuron_activation_before_labeling}
\end{figure}

To provide transparency into the learned latent representation, we visualize the average activation magnitude of each neuron across the full item dataset.
This analysis allows us to assess the relative utilization of the learned latent dimensions and verify that the sparse autoencoder does not collapse to a small subset of dominant neurons. Figure~\ref{fig:msae_overall_neuron_activation_before_labeling} presents the mean activation values of all 50 neurons sorted by activation magnitude before applying semantic labeling.
This view shows the distribution of neuron usage across the latent space.

\begin{figure}[h]
\centering
\includegraphics[width=\linewidth, trim=0 0 0 21, clip]{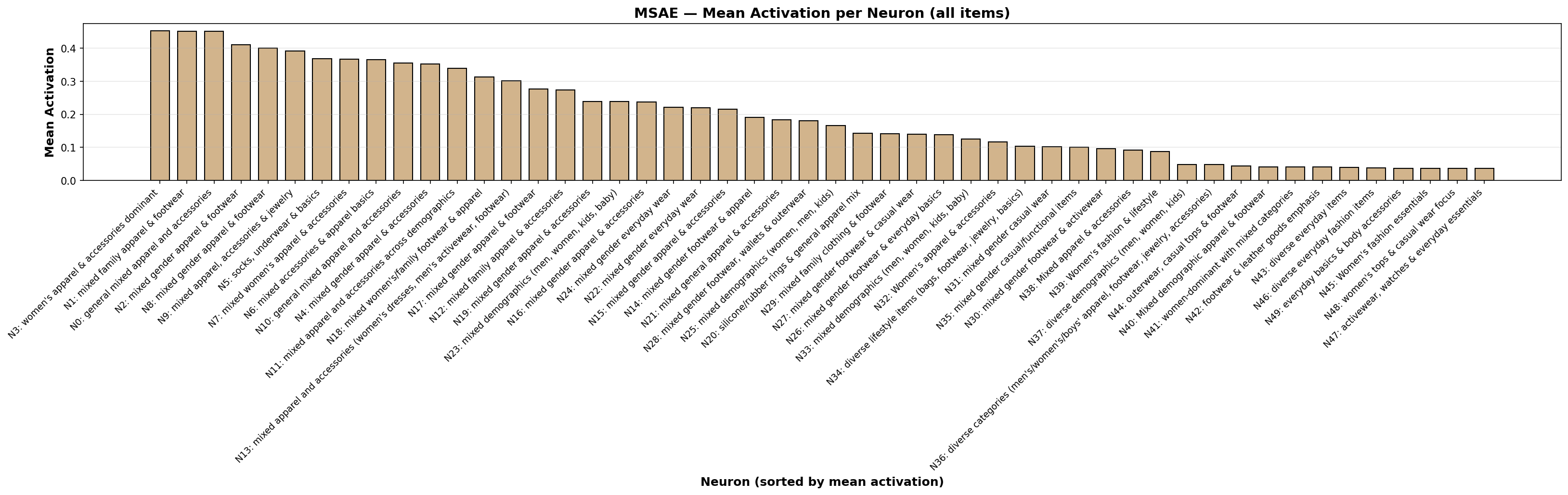}
\caption{
Mean neuron activation across the item dataset after semantic labeling using the automatic LLM-based annotation pipeline. Each neuron is associated with a high-level semantic concept inferred from the most strongly activating items, demonstrating the interpretability of the learned latent representation.
}
\label{fig:msae_overall_neuron_activation_labeled}
\end{figure}

After the automatic labeling procedure described in the previous section, neurons were associated with semantic concepts inferred from their top-activating items. Figure~\ref{fig:msae_overall_neuron_activation_labeled} presents the same activation statistics but annotated with the corresponding semantic labels produced by the LLM pipeline.

These visualizations illustrate how the latent representation organizes items according to interpretable semantic themes such as apparel categories, footwear types, accessories, and demographic segments. These activation statistics also provide a sanity check that the latent
representation is well utilized and that semantic labeling is applied to neurons that meaningfully participate in the representation.

\newpage
\begin{thebibliography}{10}

\bibitem{Ricci2011IntroRecSys}
Ricci, Francesco, Lior Rokach, and Bracha Shapira.
"Introduction to recommender systems handbook."
Recommender Systems Handbook (2010): 1--35.

\bibitem{Koren2009}
Koren, Yehuda, Robert Bell, and Chris Volinsky. "Matrix factorization techniques for recommender systems." Computer 42.8 (2009): 30-37.

\bibitem{He2017NCF}
He, Xiangnan, Lizi Liao, Hanwang Zhang, Liqiang Nie, Xia Hu, and Tat-Seng Chua.
"Neural collaborative filtering."
Proceedings of the 26th International Conference on World Wide Web. 2017.

\bibitem{Marlin2003Modeling}
Marlin, Benjamin M. "Modeling user rating profiles for collaborative filtering." Advances in neural information processing systems 16 (2003).


\bibitem{Zhang2018Explainable}
Zhang, Yongfeng, and Xu Chen. "Explainable recommendation: A survey and new perspectives." Foundations and Trends® in Information Retrieval 14.1 (2020): 1-101.

\bibitem{Abdollahi2019ExplainableMF}
Abdollahi, Behnoush, and Olfa Nasraoui. "Explainable matrix factorization for collaborative filtering." Proceedings of the 25th International Conference Companion on World Wide Web. 2016.

\bibitem{NAIS}
He, Xiangnan, Fuli Feng, Xiangnan He, Xiang Wang, Yong Li, and Tat-Seng Chua.
"NAIS: Neural attentive item similarity model for recommendation."
IEEE Transactions on Knowledge and Data Engineering 30.12 (2018): 2354--2366.

\bibitem{Burke2020FairnessRecSys}
Burke, Robin. "Multisided fairness for recommendation." arXiv preprint arXiv:1707.00093 (2017).


\bibitem{Ng2011SparseAE}
Ng, Andrew. "Sparse autoencoder." CS294A Lecture notes 72.2011 (2011): 1-19.

\bibitem{Chanin2024}
Chanin, David, James Wilken-Smith, Tomáš Dulka, Hardik Bhatnagar, Satvik Golechha, and Joseph Bloom.
"A is for absorption: Studying feature splitting and absorption in sparse autoencoders."
Advances in Neural Information Processing Systems 38 (2026): 82318--82355.

\bibitem{Bussmann2025Matryoshka}
Bussmann, Bart, Noa Nabeshima, Adam Karvonen, and Neel Nanda.
"Learning multi-level features with matryoshka sparse autoencoders."
arXiv preprint arXiv:2503.17547 (2025).


\bibitem{DoshiVelez2017Interpretability}
Doshi-Velez, Finale, and Been Kim. "Towards a rigorous science of interpretable machine learning." arXiv preprint arXiv:1702.08608 (2017).

\bibitem{Olah2020Circuits}
Olah, Chris, Nick Cammarata, Ludwig Schubert, Gabriel Goh, Michael Petrov, and Shan Carter.
"Zoom In: An Introduction to Circuits."
Distill 5.3 (2020): e00024--001.

\bibitem{Anthropic2024Scaling}
Templeton, Adly, Tom Conerly, Jonathan Marcus, Jack Lindsey, Trenton Bricken, Brian Chen, Adam Pearce, and others.
"Scaling monosemanticity: Extracting interpretable features from Claude 3 Sonnet."
Transformer Circuits Thread (2025).

\bibitem{Pach}
Pach, Mateusz, Shyamgopal Karthik, Quentin Bouniot, Serge Belongie, and Zeynep Akata.
"Sparse autoencoders learn monosemantic features in vision-language models."
Advances in Neural Information Processing Systems 38 (2026): 95706--95742.


\bibitem{WWW2026MonosemanticRecSys}
Arviv, Dor, Yehonatan Elisha, Oren Barkan, and Noam Koenigstein.
"Extracting interaction-aware monosemantic concepts in recommender systems."
Proceedings of the AAAI Conference on Artificial Intelligence 40.17 (2026): 14450--14458.

\end{thebibliography}
\end{document}